\RequirePackage{fix-cm}
\documentclass{svjour3}                     
\smartqed  

\usepackage{color}
\usepackage{graphicx}
\usepackage{amsmath}
\usepackage{amsfonts}
\usepackage{amssymb }
\usepackage{float}
\usepackage{multicol}
\usepackage{array}
\usepackage{tabularx}

\usepackage{latexsym}

 \journalname{Nonlinear Dynamics}
\makeatletter

\begin{document}
\title{Hidden chaotic attractors in fractional-order systems}


\titlerunning{Hidden chaotic attractors in fractional-order systems}

\author{Marius-F. Danca}

\institute{Marius-F. Danca \at
Department of Mathematics and Computer Science\\
Avram Iancu University, 400380 Cluj-Napoca, Romania \\
and \\
Romanian Institute for Science and Technology\\
400487 Cluj-Napoca, Romania \\
              \email{danca@rist.ro}}

\date{Received: date / Accepted: date}

\maketitle

\begin{abstract} In this paper, we present a scheme for uncovering hidden chaotic attractors in nonlinear autonomous systems of fractional order. The stability of equilibria of fractional-order systems is analyzed. The underlying initial value problem is numerically integrated with the predictor-corrector Adams-Bashforth-Moulton algorithm for fractional-order differential equations. Three examples of fractional-order systems are considered: a generalized Lorenz system, the Rabinovich-Fabrikant system and a non-smooth Chua system.

\vspace{3mm}
\textbf{keywords} Hidden attractor; Self-excited attractor; Fractional-order system; Generalized Lorenz System, Rabinovich-Fabrikant system, Non-smooth Chua system
\end{abstract}

\section{Introduction}
The concepts of \emph{self-excited }and \emph{hidden} attractors have been suggested recently by Leonov and Kuznetsov (see e.g. \cite{nick2,nick4,nick5,nick6}), which have become the subject of several works (various examples can be found in \cite{kap,spr,xxx1,xxx2,xxx3,xxx4,xxx5,xxx6}). The basins of attraction of hidden attractors do not intersect with small neighborhoods of any equilibrium points, while a basin of attraction of a self-excited attractor is associated with an unstable equilibrium. In this context, stationary points are less important for tracking hidden attractors than for the systems with self-excited attractors. Self-excited attractors can be localized (excited) by standard computational schemes, starting from a point in a neighborhood of some unstable equilibrium. On the other hand, for localization of hidden attractors it is necessary to develop special schemes. Some known classical chaotic and regular attractors (such as Lorenz, Chen, R\"{o}sler, van der Pol, Sprott systems, etc.) are self-excited attractors, which can be obtained numerically with standard algorithms, and are located in some neighborhoods of unstable equilibria (their basins of attraction touch upon unstable fixed points). Hidden attractors are important in practical applications because they can lead to unexpected dynamics and instability.

Some hidden attractors can be attractors in e.g. systems with no equilibria, with only one stable equilibrium, or with coexistence of attractors in multistable engineering systems (see e.g. \cite{xxx3,unu,yyy1,yyy2,xx0,nick}). Recently, coexisting hidden transient chaotic attractors have been found in the Rabinovich-Fabrikant system \cite{dancax}.

Uncovering all co-existing attractors and their underlying basins, when they exist, represents one of the major difficulties in locating hidden attractors. An analog of the famous 16th Hilbert problem (on the number and mutual dispositions of minimal chaotic attractors in the polynomial systems) is formulated in \cite{x0x0}.

Hidden attractors can be regular or chaotic. In this work, we are concerned with hidden chaotic attractors. While in the above mentioned references, hidden attractors have been found for continuous-time or discrete-time systems of integer order, in this paper we present for the first time examples of hidden attractors of three-dimensional continuous-time systems of fractional order, including a generalized Lorenz system, the Rabinovich-Fabrikant (RF) system and a non-smooth Chua system.

Hidden periodic oscillations and hidden chaotic attractors have been studied e.g. in phase-locked loop \cite{xx1}, drilling systems \cite{xx3}, DC-DC convertors \cite{xx5} or aircraft control system \cite{xx7}.

On the other hand, fractional-order systems are dynamical systems described by using fractional-order derivative and integral operators, and are studied by more and more people with growing interest. A large number of physical systems can be better modeled by means of fractional-order models \cite{sab}. Also, systems of fractional order can be found in economy \cite{scal}, bioengineering \cite{mag}, mechanics \cite{ata} etc.
Actually, real objects or phenomena such as dielectric polarization, viscoelastic systems, percolation, polymer modeling, ultra-slow processes, electromagnetic waves, evolution of complex systems, secure communication, chaotic dynamics etc. are generally of fractional order (see e.g. \cite{bag,ous,pod2,las,kus,mkai,md,moha}).\footnote{Note that even fractional-order dynamics allow to describe a real object more accurately than classical ``integer-order'' dynamics, as proved recently for the existence of stable cycles in systems of fractional order to be impossible \cite{tava4,eva}.}

Therefore, studying hidden chaotic attractors in systems of fractional order represents a good opportunity to deepen the new exciting and still less-explored subject of importance.

This paper is organized as follows: In Section 2, basic notions related to the stability of systems of fractional order, required to verify the attractors hiddenness characteristic and the numerical integration, are presented. Section 3 considers the hidden attractors of a generalized Lorenz system, the Rabinovich-Fabrikant system and a non-smooth Chua system. Finally a conclusion ends the paper.

\section{Stability and discretization of fractional-order systems }

The considered dynamical systems are modeled by the following fractional-order initial values problem (IVP):
\begin{equation}\label{eq:General_IVP_FDE}
	\frac{d^{q}}{dt^{q}} x(t) = f(x(t)), \quad x(0)=x_0, \quad t\in I=[0,T],
\end{equation}
\noindent where  $x:I \to \mathbb{R}^n$, $f:\mathbb{R}^{n}\rightarrow\mathbb{R}^{n}$ is a continuous nonlinear function and $q\in(0,1)$ represents the commensurate order of the derivatives. For basic knowledge on fractional calculus, one may refer to \cite{ous,pod2,Podlubny1999,Diethelm2010,KilbasSrivastavaTrujillo2006,bale2}. In this work, we consider the fractional derivative operator $d^{q}/dt^{q}$, with $q<1$, to be \emph{Caputo}'s derivative with starting point $t_0=0$ defined by \cite{Podlubny1999}
\begin{equation}\label{eq:CaputoDefinition}
	\frac{d^{q}}{dt^{q}} x(t) =
	\frac{1}{\Gamma(1-q)} \int_{0}^{t} \bigl(t-s\bigr)^{-q} x'(s) ds ,
\end{equation}
where $\Gamma$ is the Euler gamma function. The use of Caputo's definition allows coupling the fractional differential equations with initial conditions in a classical form and avoids the expression of initial conditions with fractional derivatives. Note that coupling differential equations with classical initial conditions of Cauchy type not only has a clearly interpretable physical meaning but also can be measured to properly initializing simulations\footnote{Recently, based on philosophical arguments rather than a mathematical point of view, some researchers questioned the appropriateness of using initial conditions of the classical form in the Caputo derivative \cite{hart}. However, it should be emphasized that, in practical (physical) problems, physically interpretable initial conditions are necessary and Caputo's derivative is a fully justified tool \cite{kai2}.} (see \cite{Diethelm2010} for more insights on this topic and relationship to the case of $q>1$).

The right-hand side of the IVP (\ref{eq:General_IVP_FDE}) in the considered examples are Lipschitz functions, and the numerical method used in this work to integrate system (\ref{eq:General_IVP_FDE}) is the Adams-Bashforth-Moulton predictor-corrector algorithm \cite{kai}.
Specifically, the algorithm works by introducing a discretization with grid points $t_i=hi, i=0,1,...$, and a preassigned step size $h$. For some fractional-order $q$, and $i=0,1,2,...$, it first computes a preliminary approximation (predictor) denoted as $x^P_{i+1}$ for $x(t_{i+1})$ using the formula

\[
x_{i+1}^P=\sum_{j=0}^{\lceil q\rceil-1} x_0^{(j)}\frac{t^j_{i+1}}{j!}+\frac{1}{\Gamma(q)}\sum_{j=0}^ib_{j,i+1}f(x_j),
\]

\noindent with

\[
b_{j,i+1}=\frac{h^q}{q}\big((i+1-j)^q-(i-j)^q\big),
\]

\noindent and then calculates the corrector value $x_{i+1}$ by

\[
x_{i+1}=\sum_{j=0}^{\lceil q\rceil-1}x_0^{(j)}\frac{t_{i+1}^j}{j!}+\frac{h^q}{\Gamma(q+2)}\Bigg(\sum_{j=0}^ia_{j,i+1}f(x_j)+f(x_{i+1}^P)\Bigg),
\]

\noindent where

\[
a_{j,i+1}=\left\{
\begin{array}{ll}
i^{q+1}-(i-q)(i+1)^{q}, & j=0, \\
(i-j+2)^{q+1}+(i-j)^{q+1}-2(i-j+1)^{q+1}, & 1\leq j\leq i, \\
1, & j=i+1.%
\end{array}%
\right.
\]

To define the stability of equilibria of fractional-order systems (required by the procedure to find hidden attractors), consider some equilibrium $X^*$ and the Jacobian $J=\frac{\partial f}{\partial x}|_{x=X^*}$ evaluated at $X^*$. Denote by $\Lambda=\{\lambda_1,\lambda_2,...,\lambda_n\}$ the eigen-spectrum and let the minimum of all arguments of the eigenvalues be $\alpha_{min}=min\{|\alpha_i|\}$, $i=1,2,...,n$. Then, a stability theorem \cite{tava,tava3} can be stated in the following practical form:

\begin{theorem}

$X^*$ is asymptotically stable if and only if the instability measure

\begin{equation}\label{cond}
\iota=q-2\alpha_{min}/\pi
\end{equation}

\noindent is strictly negative.
\end{theorem}

If $\iota\leq 0$ and the critical eigenvalues satisfying $\iota=0$ have the geometric multiplicity one, then $X^*$ is stable.\footnote{The geometric multiplicity represents the dimension of the
eigenspace of the corresponding eigenvalues.}

\begin{remark}
If $\iota$ is positive, then $X^*$ is unstable and the system may exhibit chaotic behavior.
\end{remark}

\section{Hidden chaotic attractors}

From a computational point of view, self-excited attractors and hidden attractors are defined as follows:

\begin{definition}\cite{nick2,nick4,nick5}
\label{def}
An attractor is called a self-excited
attractor if its basin of attraction intersects with any
open neighborhood of an equilibrium,
otherwise it is called a hidden attractor.
\end{definition}

Self-excited attractors can be obtained numerically with standard computational schemes, in which after transients being eliminated, the trajectories starting from neighborhoods of unstable equilibria are attracted by the attractor. In contrast, the basin of attraction for a hidden attractor is not connected with any equilibrium. Therefore, for the numerical localization of hidden attractors it is necessary to develop special analytical-numerical algorithms (see e.g. \cite{nick2} and \cite{xx0}). The first stage in the localization requires a harmonic linearization procedure,
which allows one to modify the system such that its linear part has a periodic solution\footnote{ In many cases, one can simplify this procedure and consider instead a path in the space of parameters, such that the starting point of the path corresponds to a self-excited attractor.}. Next step is to modify the nonlinearity by introducing a small parameter. This parameter must be small enough in order to generate a periodic solution, which will be the first step of the multi-step numerical continuation procedure: construct a sequence of similar systems such that for the first (starting) system the initial point for numerical computation of oscillating solution (starting oscillation) can be obtained analytically (e.g, it is often possible to consider the starting system
with self-excited starting oscillation). Then, the transformation of this starting oscillation is followed numerically
in passing from one system to another and the last system will correspond to the hidden attractor.

Summarizing, to obtain a hidden attractor it is necessary to first verify that it is characterized by Definition \ref{def}: Supposing that the system admits stable and unstable equilibria (as in our considered examples). This means one should verify numerically that trajectories starting from vicinities of unstable equilibria either are attracted by stable equilibria or tend to infinity. The next step is to visualize the hidden attractor, for example by following the procedure described in \cite{nick2}. However, for such as the examples considered in this work, the try-and-error method can be utilized also for plotting the hidden attractor.

In order to obtain significant (stronger) chaotic behaviors, the fractional-order $q$ for the considered systems is chosen to be relatively high (close to $1$).

\subsection{Hidden chaotic attractor of a generalized Lorenz system of fractional order}

For each considered system, several numerical experiments to generate sets of 100 trajectories starting from each unstable equilibrium have been done. However, for the image clarity, only representative trajectories are presented here.

The generalized Lorenz system of fractional order is a fractional variant of the generalized Lorenz system of integer order \cite{xx0,leo2} and is obtained from a Rabinovich system \cite{rab1,rab2} as follows:

\begin{equation}
\label{lor}
\begin{array}{l}
{x}_{1}^q=-\sigma (x_1-x_2)-ax_2x_3, \\
{x}_{2}^q=rx_1-x_2-x_1x_3,\\
{x}_{3}^q=-x_3+x_1x_2,
\end{array}%
\end{equation}

\noindent where $\sigma=-ar$ and $a<0$. For $a=0$, the system (\ref{lor}) coincides with the classical Lorenz system.

The system \eqref{lor} can be used to describe: the convective fluid motion inside rotating ellipsoid \cite{rus1}, the rotation of rigid body in viscous fluid \cite{rus2}, the gyrostat dynamics \cite{rus3}, the convection of horizontal layer of fluid making harmonic oscillations, or the model of Kolmogorov's flow \cite{rus4}.

Due to the symmetry

\begin{equation}\label{sim}
T(x_1,x_2,x_3)\rightarrow (-x_1,-x_2,x_3),
\end{equation}

\noindent under transformation $T$, each trajectory has its symmetrical ("twin") trajectory with respect to the $x_3$-axis.

As mentioned in \cite{xx0}, for $r<1$, there exists a unique equilibrium $X^*_0=(0,0,0)$, while for $r>1$ there exist three equilibria: $X^*_0$ and

\[
X_{1,2}^*=(\pm x^*,\pm y^*,z^*),
\]

\noindent with

\[
x^*=\frac{\sigma \sqrt{\xi}}{\sigma+a \xi},\quad y^*=\sqrt{\xi},\quad z^*=\frac{\sigma\xi}{\sigma+a\xi},
\]

\noindent where

\[
\xi=\frac{\sigma}{2a^2}\big[ a(r-2)-\sigma+\sqrt{(\sigma-ar)^2+4a\sigma}\big].
\]

Let $r = 6.8$ and $a = -0.5$. Then, equilibria $X_{1,2}^*$ are given by

\[
X_{1,2}^*=(\pm 3.476,\pm 1.807,6.280),
\]

\noindent and the integer-order system (\ref{lor}) presents a hidden attractor \cite{xx0}.

Here, we focus on the existence of a hidden attractor for the case of this fractional-order system. The Jacobian matrix is

\[
   J=
  \begin{bmatrix}
   -\sigma & \sigma-ax_3 & -ax_2 \\ r-x_3& -1 & -x_1\\x_2& x_1 & -1\end{bmatrix}.
\]

Consider the equilibrium $X_0^*$. The Jacobian evaluated at this point has the eigen-spectrum $\Lambda=\{2.5576, -1,-7.5576\}$ with arguments: $\alpha_1=0$ and $\alpha_{2,3}=\pm\pi$ and $\alpha_{min}=0$. In this case, the instability measure (\ref{cond}) is $\iota=q-2\alpha_{min}\pi/2=q>0$ for all $q\in(0,1)$, so the equilibrium $X_0^*$ is unstable.

Consider the equilibrium $X_1^*$ (due to the system symmetry, $T$, $X_2^*$ behaves similarly). The eigen-spectrum is $\Lambda=\{-5.9570,- 0.0215-3.6026i,- 0.0215+3.6026i\}$ with arguments $\alpha_1=\pi$, $\alpha_{2,3}=\pm 1.5768$ and $\alpha_{min}=1.5768$. Since the instability measure is $\iota=q-2\alpha_{min}/\pi=q-1.0038<$ for all $q\in(0,1)$, the equilibria $X_{1,2}^*$ are asymptotically stable for all $q\in(0,1)$, so they are saddle points.

Note that the stability of equilibria $X_{0,1,2}^*$ doesn't change for any values of $q<1$, which is similar to the integer-order case.

\begin{figure}
\begin{center}
\includegraphics[scale=0.5]{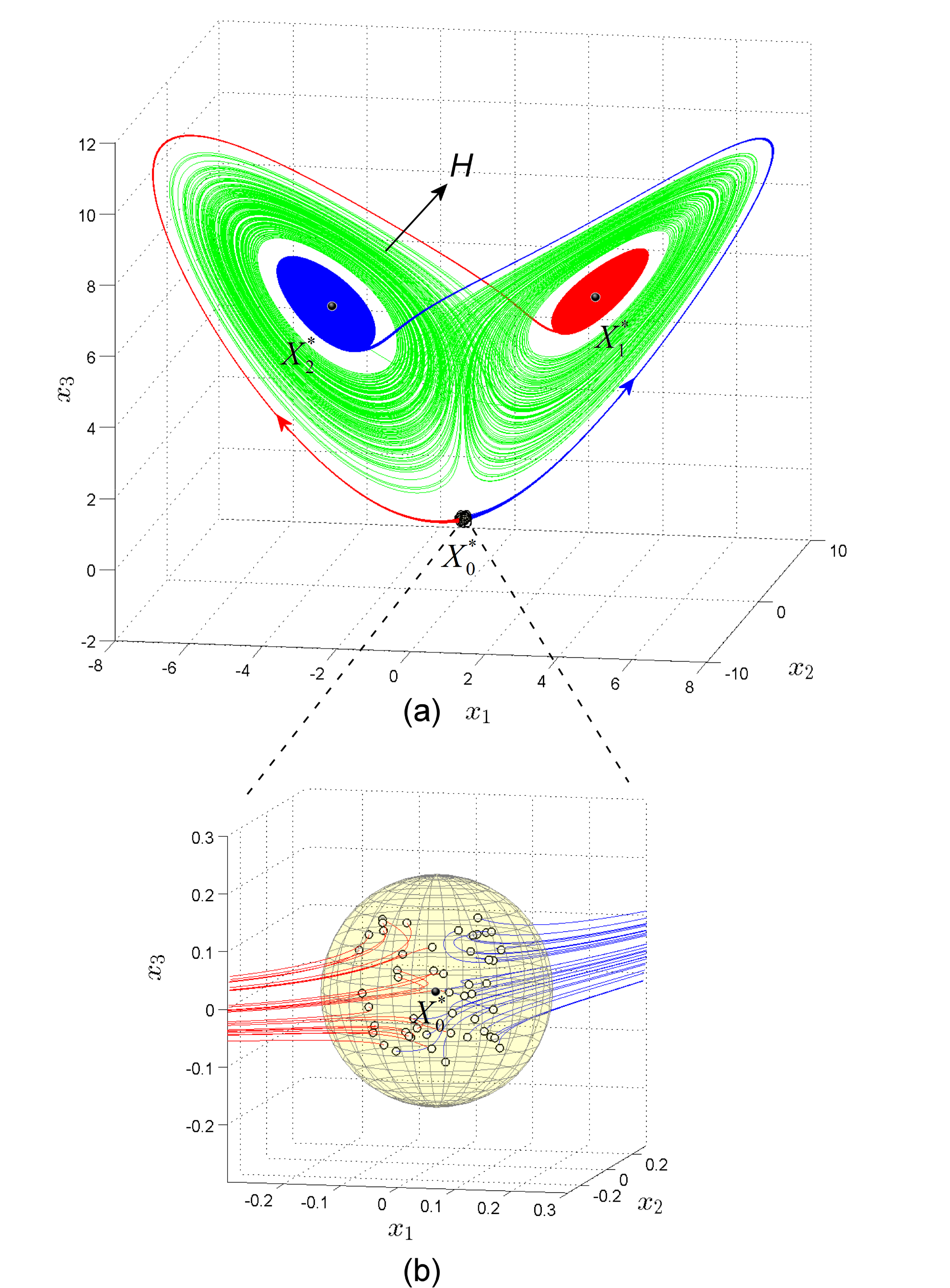}
\caption{(a) Hidden attractor of the generalized Lorenz system of fractional-order (green). The red and blue trajectories, starting from a small vicinity of the unstable equilibrium $X_0^*$, are attracted by the stable equilibria $X_{1,2}^*$. (b) The detailed image shows how the considered 50 trajectories tend either to $X_1^*$ (red), or to $X_2^*$ (blue).}
\label{fig1}
\end{center}
\end{figure}

Consider $q=0.995$, a value for which the system exhibits chaotic behavior.
In order to verify that the generalized Lorenz system (\ref{lor}) has a hidden attractor, we have to verify that there exist small neighborhoods of the unstable equilibrium $X_0^*$, orbit from which are all attracted by the stable equilibria $X_{1,2}^*$ (Fig. \ref{fig1}). As can be seen in Fig. \ref{fig1} (a), trajectories exiting from a small vicinity of $X_0^*$ either tend to $X_1^*$ (red plot), or to $X_2^*$ (blue plot). In the detail in Fig. \ref{fig1} (b), the considered 50 trajectories starting from a vicinity of ray $\delta=0.3$ centered at $X_0^*$ show how they are attracted either by $X_1^*$, or $X_2^*$. The hidden chaotic attractor $H$ is colored in green.

\subsection{Hidden chaotic attractor of the Rabinovich-Fabrikant system}

The fractional-order RF system \cite{dan,dan2} is modeled by

\begin{equation}
\label{rf}
\begin{array}{l}
{x}_{1}^q=x_{2}\left( x_{3}-1+x_{1}^{2}\right) +ax_{1}, \\
{x}_{2}^q=x_{1}\left( 3x_{3}+1-x_{1}^{2}\right) +ax_{2}, \\
{x}_{3}^q=-2x_{3}\left( b+x_{1}x_{2}\right),
\end{array}%
\end{equation}

\noindent with $a>0$ and $b$ being the bifurcation parameter. The system, initially designed as a physical system, describes the stochasticity arising from the modulation instability in a dissipative medium. However, as revealed numerically in \cite{dan} and \cite{dan2}, the system of integer order presents unusual and extremely rich dynamics, including multistability, an important ingredient for potential existence of hidden attractor.

The equilibria are $X_0^*$ and

\begin{equation}
\label{eq}
\begin{array}{l}
X_{1,2}^{\ast }\left( \mp x^*_{1,2},\pm y_{1,2}^*,z_{1,2}^*\right),\quad
X_{3,4}^{\ast }\left( \mp x_{3,4}^*,\pm y_{3,4}^*,z_{3,4}^*\right) ,
\end{array}%
\end{equation}

\noindent where
\[
x^*_{1,2}= \sqrt{\dfrac{bR_1+2b}{4b-3a}},\quad y^*_{1,2}=\sqrt{b\dfrac{4b-3a}{R_1+2}},\quad z^*_{1,2}=\dfrac{aR_1+R_2}{\left(4b-3a\right) R_1+8b-6a}\\,
\]

\noindent and

\[
x_{3,4}^*=\sqrt{\dfrac{bR_1-2b}{3a-4b}},\quad y_{3,4}^*=\sqrt{b\dfrac{4b-3a}{2-R_1}},\quad z_{3,4}^*=\dfrac{aR_1-R_2}{\left(
4b-3a\right) R_1-8b+6a},
\]

\noindent with $R_1=\sqrt{3a^{2}-4ab+4}$ and $R_2=4ab^{2}-7a^{2}b+3a^{3}+2a$.

Let $a=0.1$ and $b=0.2876$. Then, the equilibria $X_{1,2,3,4}^*$ are

\[
X_{1,2}^*=(\mp1.1600,\pm0.2479, 0.1223),\quad X_{3,4}^*=(\mp0.0850,\pm3.3827, 0.9953).
\]

It is easy to see that the system exhibits the same symmetry $T$ defined in (\ref{sim}), as for the case of the generalized Lorenz system (\ref{lor}). Therefore, we can consider the stability of $X_1^*$ and $X_3^*$ only.

As required by Definition \ref{def}, we have to determine the stability of all equilibria. The Jacobian is

\[
   J=
  \begin{bmatrix}
  2x_1x_2+a& x_1^2+x_3-1& x_2\\-3x_1^2+3x_3+1& a& 3x_1\\-2x_2x_3& -2x_1x_3& -2(x_1x_2+b)
\end{bmatrix}.
\]

Consider, first, the equilibrium $X_0^*$. The spectrum of eigenvalues is $\Lambda=\{-0.5752,0.1-i,0.1+i\}$, with arguments: $\alpha_1=\pi$, $\alpha_{2,3}=\pm1.4711$, $\alpha_{min}=1.4711$. The instability measure is $\iota=q-2\alpha_{min}/\pi=q-0.9365>0$ for $q>0.9365$. Therefore, $X_0^*$ is an unstable focus-saddle only for $q>0.9365$, while for smaller values of $q$ the equilibrium $X_0^*$ is stable.

The eigen-spectrum of the equilibrium $X_1^*$ is $\Lambda=\{-0.2562,-0.0595-1.4731i,-0.0595+1.4731i\}$ with arguments: $\alpha_1=\pi$, $\alpha_{2,3}=\pm 1.6112$, $\alpha_{min}=1.6112$ and
$\iota=q-1.0257<0$ for all $q\in(0,1)$. Therefore, $X_{1,2}^*$ are stable for all $q\in(0,1)$.

Finally, the spectrum of the equilibrium $X_{3}^*$ is $\Lambda=\{0.1981,-0.2866-4.7743i,-0.2866+4.7743i\}$ with arguments: $\alpha_1=0$, $\alpha_{2,3}=\pm1.6308$ and $\alpha_{min}=0$. In this case, $\iota=q>0$ and, therefore, $X_{3,4}^*$ are unstable for all $q\in(0,1)$.

Taking account on the instability of the equilibrium $X_0^*$, we consider $q=0.998$.

\begin{figure}
\begin{center}
\includegraphics[scale=0.45]{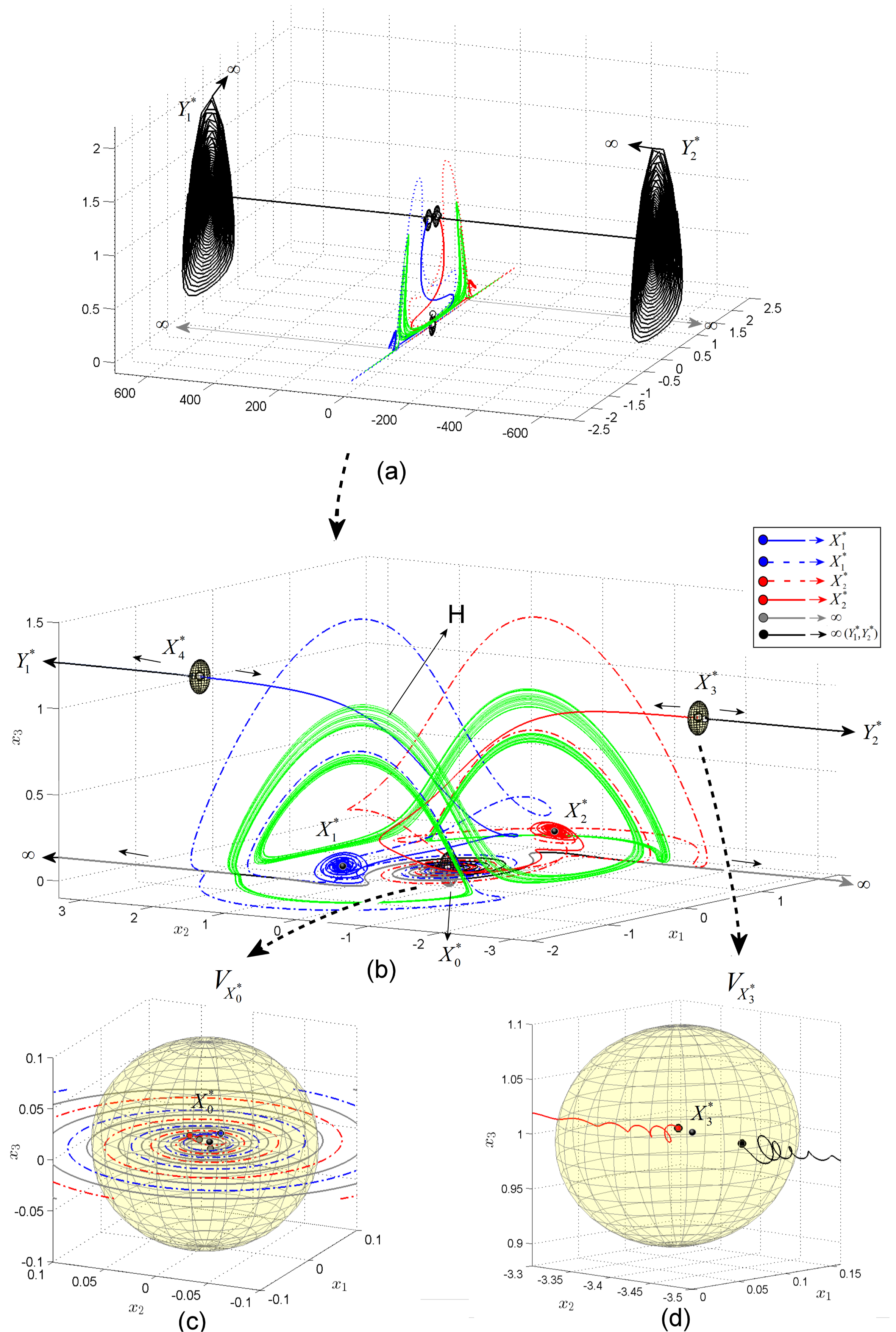}
\caption{(a) Hidden attractor $H$ (green) and ``virtual'' repelling saddles $Y_{1,2}^*$ (black) of the generalized RF system of fractional-order. (b) The detailed image presents the hidden attractor, equilibria, and the trajectories starting from vicinities of unstable equilibria $X_{0,3,4}^*$. (c) Detailed region of the unstable equilibrium $X_0^*$ revealing four trajectories, which tend either to infinity (grey) or to equilibria $X_{1,2}^*$ (dotted blue and dotted red respectively). (d) Detailed region of the unstable equilibrium $X_3^*$ revealing two trajectories, which tend either to infinity (black) or to equilibria $X_{1,2}^*$ (blue and red respectively).}
\label{fig2}
\end{center}
\end{figure}

For the considered parameter values and fractional order, beside the hidden attractor $H$ (green plot in Fig. \ref{fig2} (a)) the system presents unusual behavior such as ``unbounded self-excited attractors'' $Y_{1,2}^*$ (Fig. \ref{fig2} (a)) which, due to the resemblance with saddles, are called ``virtual'' repelling saddles \cite{dan2}. As shown in \cite{dan2}, these saddles-like attractors exist for relatively large domains of parameters $a$ and $b$.

To check that the chaotic attractor $H$ is hidden (see the detailed image in Fig.\ref{fig2} (b), where the equilibria $X_{0,1,2,3,4}^*$ beside $H$ can be viewed), we have to verify numerically that all trajectories starting from small vicinities of all unstable equilibria ($X_0^*$ and $X_{3,4}$) either diverge to infinity or are attracted by the stable equilibria $X_{1,2}^*$. As can be seen in Fig. \ref{fig2} (b) and Fig. \ref{fig2} (c), all trajectories starting from the vicinity of $X_0^*$ either diverge to infinity (grey plot), or converge to the stable equilibria $X_{1,2}$ (dotted blue and dotted red plot, respectively). From Fig. \ref{fig2} (b) and Fig. \ref{fig2} (c), one can see that the trajectories starting from a vicinity of $X_3^*$ ($X_4^*$ leads to similar behavior) tend either to the ``virtual'' repelling saddle $Y^*$, or are attracted by $X_{1,2}^*$ (blue plot and red plot respectively). The sizes of the vicinities in this case have been chosen as $\delta=0.1$, and for the sake of image clarity, only representative trajectories are plotted.

\subsection{Hidden chaotic attractor of a non-smooth Chua system of fractional order}
Consider the non-smooth Chua system modeled by \cite{nickx,kuzbb}

\begin{equation}
\label{rf}
\begin{array}{l}
{x}_{1}^q=\alpha(x_2-x_1-m_1x_1-\psi(x_1)),\\
{x}_{2}^q=x_1-x_2+x_3, \\
{x}_{3}^q=-(\beta x_2+\gamma x_3),\quad
\end{array}%
\end{equation}

\noindent with

\[
\psi(x_1)=(m_0-m_1)sat(x_1),
\]

\noindent and

\begin{equation}
\begin{split}\nonumber
m_0=-0.1768, \\
m_1=-1.1468, \\
\alpha=8.4562, \\
\beta=12.0732, \\
\gamma=0.0052.
\end{split}
\end{equation}

After almost 30 years of its first investigation of Chua's circuits of integer or fractional order, only self-excited attractors have been found. However, later it was shown (see e.g. \cite{nickx}) that Chua's circuits of integer order has hidden chaotic attractors, with a positive largest Lyapunov exponent.

The system is continuous non-smooth because of the function $\psi$. But, its right-hand side is locally Lipschitz from the absolute value operator in $\psi(x_1)$, so that the ABM method can be applied \cite{Diethelm2010}.

Equilibria are: $X_0^*=(0,0,0)$ for $|x|<1$ and $X_{1,2}^*=(\pm6.5883, \pm 0.0029 \mp 6.5855)$ for $|x|>1$, and the Jacobian is

\[
   J=
  \begin{bmatrix}
  -\alpha-\alpha m_1& \alpha& 0\\ 1& -1& 1\\0&-\beta&-\gamma
\end{bmatrix},\quad \text{for}\quad |x|>1,
\]

\noindent and

\[
   J=
  \begin{bmatrix}
  -\alpha-\alpha m_1-m_0+m_1& \alpha& 0\\ 1& -1& 1\\0&-\beta&-\gamma
\end{bmatrix},\quad \text{for}\quad |x|<1.
\]

The eigen-spectrum of $X_0^*$ is $\Lambda=\{-7,9587, -0.0038\pm 3.2494i\}$ and $\iota=q-1.0008<0$ for all $q\in(0,1)$. Therefore, $X_0^*$ is asymptotically stable.

For $X_{1,2}^*$, $\Lambda=\{2.2193,-0.9916\pm 2.4068i\}$ and $\iota=q>0$ for all $q\in(0,1)$ and the equilibria are unstable.

\begin{figure}
\begin{center}
\includegraphics[scale=0.4]{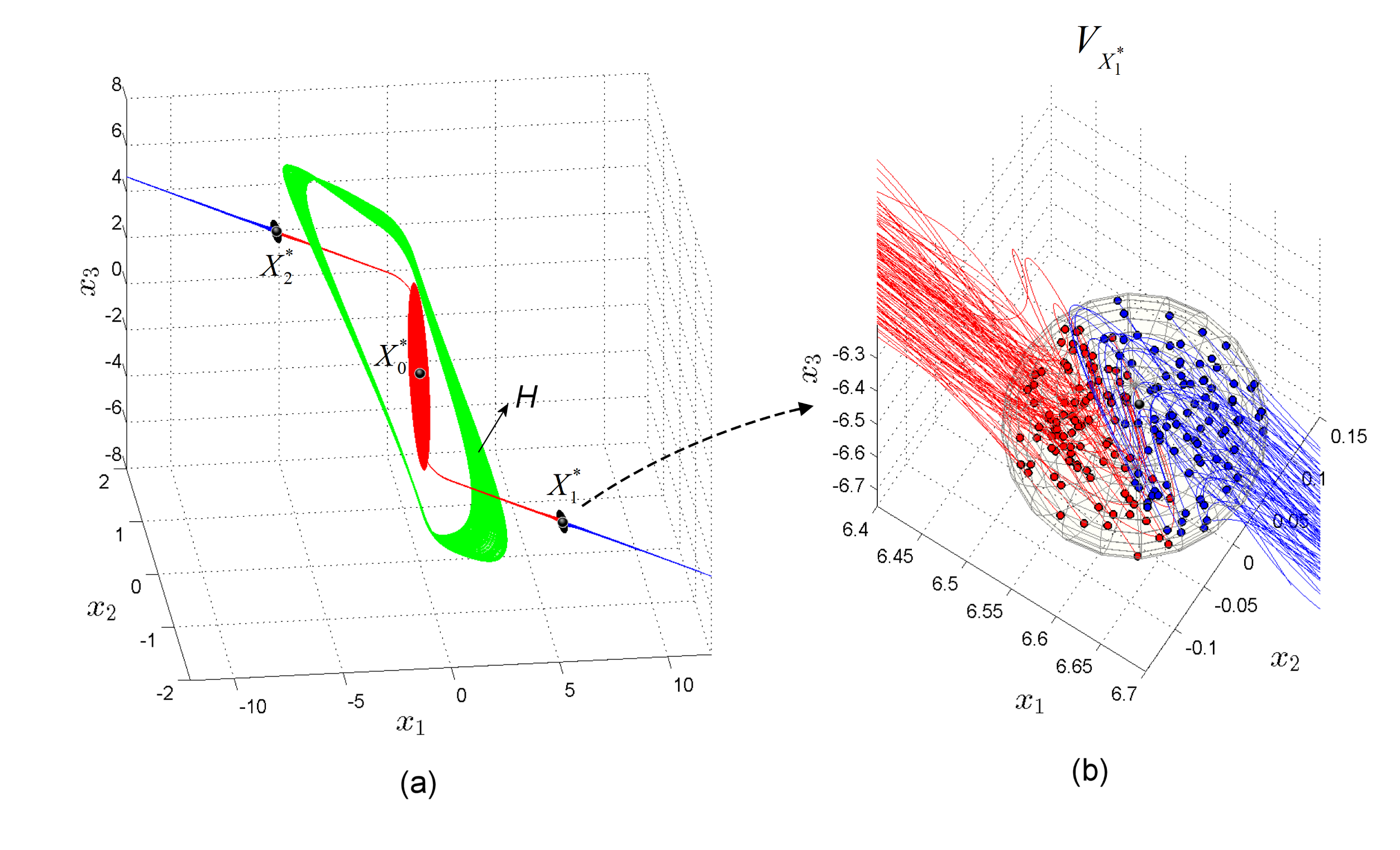}
\caption{Hidden attractor (green) of the non-smooth Chua system. (b) The detailed image presents $200$ trajectories starting from a vicinity of the equilibrium $X_1^*$ tending either to $X_0^*$ (red) or to infinity (blue).}
\label{fig3}
\end{center}
\end{figure}

For $q=0.9998$, the numerical results are presented in Fig. \ref{fig3} (a), while the detail in Fig. \ref{fig3} (b) reveals the tendency of all trajectories starting from a small vicinity of unstable equilibria (of size $\delta=0.01$) tend either to the stable equilibrium $X_0^*$ (red plot) or diverge to infinity (blue plot).

\section{Conclusion}

Our new results show that smooth or non-smooth fractional-order three-dimensional systems of commensurates order can exhibit hidden chaotic attractors. The presented approach can be successfully
implemented also for other systems of fractional order. Similarly, one can search hidden attractors for fractional-order systems of incommensurate orders.
In addition, we have revealed that the unusual and extremely rich dynamics (``virtual'' saddles-like) of the RF system, found previously for the integer-order system, persist for the corresponding fractional-order variant.

\vspace{2mm}
\textbf{Acknowledgments} MF Danca is supported by Tehnic B SRL.

\newpage{\pagestyle{empty}\cleardoublepage}


\begin{thebibliography}{100}



\bibitem{nick2} Leonov, G.A., Kuznetsov, N.V. Vagaitsev, V.I.: Hidden attractor in smooth Chua systems. Physica D \textbf{241}(18), 1482-1486 (2012)

	\bibitem{nick4} Leonov, G., Kuznetsov, N., Mokaev, T.: Homoclinic orbits, and self-excited and hidden attractors in a Lorenz-like system describing convective fluid motion . Eur. Phy. J.-Spec. Top. \textbf{224}(8), 1421-1458 (2015)

	\bibitem{nick5} Leonov, G.A., Kuznetsov, N.V.: Hidden attractors in dynamical systems. From hidden oscillations in Hilbert-Kolmogorov, Aizerman, and Kalman problems to hidden chaotic attractor in Chua circuits. Int. J. Bifurcat. Chaos \textbf{23} 1330002 (2013)

	\bibitem{nick6} Kuznetsov, N.V.: Hidden Attractors in Fundamental Problems and Engineering Models: A Short Survey. Chapter in Lect. Notes Electr. En. \textbf{371} 13 (2016)

	\bibitem{kap} Dudkowski, D., Prasad, A., Kapitaniak, T.: Perpetual points and hidden attractors in dynamical systems. Phys. Lett. A \textbf{379}(40-41), 2591-2596 (2015)

	\bibitem{spr} Jafari, S., Sprott, J.C., Nazarimehr, F.: Recent new examples of hidden attractors, Eur. Phy. J.-Spec. Top. \textbf{224}(8), 1469- 1476 (2015)

	\bibitem{xxx1} Shahzad, M., Pham, V.T., Ahmad, M., Jafari, S., Hadaeghi, F.: Synchronization and circuit design of a chaotic system with coexisting hidden attractors. Eur. Phy. J.-Spec. Top. \textbf{224}(8) 1637- 1652 (2015)

	\bibitem{xxx2} Brezetskyi, S., Dudkowski, D., Kapitaniak, T.: Rare and hidden attractors in Van der Pol-Duffing oscillators. Eur. Phy. J.-Spec. Top. \textbf{224}(8), 1459- 1467 (2015)

	\bibitem{xxx3} Cafagna, D., Grassi, G.: Fractional-order systems without equilibria: The first example of hyperchaos and its application to synchronization. Chinese Phys. B \textbf{24}(8) 080502 (2015)

	\bibitem{xxx4} Zhusubaliyev, Z.T., Mosekilde, E.: Multistability and hidden attractors in a multilevel DC/DC converter. Math. Comput. Simul. \textbf{109}, 32-45 (2015)

	\bibitem{xxx5} Chen, M., Li, M., Yu, Q., Bao, B., Xu, Q., Wang, J.: Dynamics of self-excited attractors and hidden attractors in generalized memristor-based Chua\textsc{\char13}s circuit. Nonlinear Dynam. \textbf{81}(1-2), 215-226 (2015)

	\bibitem{xxx6} Wang, Z., Sun, W., Wei, Z., Zhang, S.: Dynamics and delayed feedback control for a 3D jerk system with hidden attractor. Nonlinear Dynam. \textbf{82}(1-2) 577-588 (2015)

	\bibitem{unu} Wei, Z., Wang, R., Liu, A.: A new finding of the existence of hidden hyperchaotic attractors with no equilibria. Math. Comput. Simulat. \textbf{100}, 13-23 (2014)

	\bibitem{yyy1} Sprott, J.C., Jafari, S., Pham, V.-T., Hosseini, Z.S.: A chaotic system with a single unstable node. Phys. Lett. A \textbf{379}(36), 2030-2036 (2015)

	\bibitem{yyy2} Heath, W.P., Carrasco, J., de la Sen, M.: Second-order counterexamples to the discrete-time Kalman conjecture. Automatica \textbf{60}, 140-144 (2015)

	\bibitem{xx0} Kuznetsov, N.V., Leonov, G.A., Mokaev, T.N.: Hidden attractor in the Rabinovich system, arXiv:1504.04723v1 (2015) http://arxiv.org/pdf/1504.04723v1.pdf.

	\bibitem{nick} Kuznetsov, N.V., Leonov, G.A.: Hidden attractors in dynamical systems: systems with no equilibria, multistability and coexisting attractors. IFAC Proceedings Volumes, 19th IFAC World Congress, \textbf{47}(3), 5445-5454 (2014)

	\bibitem{dancax} M.-F. Danca, Hidden transient chaotic attractors of Rabinovich-Fabrikant system. Nonlinear Dynam. \textbf{86}(2), 1263–1270 (2016)

	\bibitem{x0x0} Leonov G.A., Kuznetsov N.V.: On differences and similarities in the analysis of Lorenz, Chen, and Lu systems. Appl. Math. Comput. \textbf{256} 334-343 (2015)
\bibitem{xx1}Kuznetsov, N.V., Kuznetsova, O.A, Leonov, G.A., Neittaanmuaki, P., Yuldashev, M.V., Yuldashev, R.V.: Limitations of the classical phase-locked loop analysis. Proceeding-IEEE International Symposion on Circuits and Systems, pp. 533-536, art. no. 7168688 (2015)
     \bibitem{xx3}Leonov, G.A., Kuznetsov, N.V., Kiseleva, M.A.,  Solovyeva, E.P., Zaretskiy, A.M.: Hidden
oscillations  in  mathematical  model  of  drilling  system  actuated  by  induction  motor  with  a
wound rotor. Nonlinear Dynam. \textbf{77}(1-2), 277–288  (2014)
\bibitem{xx5}Zhusubaliyev, Z.T., Mosekilde, E.: Multistability and hidden attractors in a multilevel DC/DC converter. Math. Comput. Simulat. \textbf{109}, 32–45 (2015)
\bibitem{xx7}Andrievsky. B.R., Kuznetsov, N.V., Leonov, G.A., Pogromsky, A.Yu.: Hidden oscillations in aircraft flight control system with input saturation. IFAC Proceedings Volumes \textbf{46}(12), 75-79 (2013)

	\bibitem{sab} Sabatier, J., Agrawal,  O.P., Machado, Tenreiro, J.A.: \emph{Advances in fractional calculus; Theoretical developments and spplications, }Physics and Engineering Series, (Springer, Berlin, 2007).

	\bibitem{scal} Scalas, E., Gorenflo, R., Mainardi, F.: Fractional calculus and continuous-time finance. Physica A \textbf{285}(1-4), 376-384 (2000)

	\bibitem{mag} Magin, R.L.: \emph{Fractional Calculus in Bioengineering, }(Begell House, Danbury, CT, 2006).

	\bibitem{ata} Atanackovi\'{c}, T.M.: On a distributed derivative model of a viscoelastic body. C. R. Mecanique \textbf{331}(10), 687-692 (2003)

	\bibitem{bag} Bagley, R.L., Calico, R.A.: Fractional order state equations for the control of viscoelasticallydamped structures. J. Guid. Control Dyn. \textbf{14}(2), 304-311 (1991)

	\bibitem{ous} Oustaloup, A.: \emph{La derivation non entiere: Theorie, synthese et applications, }(Hermes, Paris, 1995).

	\bibitem{pod2} Podlubny, I., Petr\'{a}\u{s}, I., Vinagre, B.M., O'Leary, P., Dorc\'{a}k, K.: Analogue realizations of fractional-order controllers. Nonlinear Dynam. \textbf{29}(1) 281-296 (2002)

	\bibitem{las} Laskin, N.: Fractional market dynamics. Physica A \textbf{287}(3-4), 482-492 (2000)

	\bibitem{kus} Kusnezov, D., Bulgac, A., Dang, G.D.: Quantum L\'{e}vy Processes and Fractional Kinetics. Phys. Rev. Lett. \textbf{82}(6) 1136-1139 (1999)

	\bibitem{mkai} Danca, M.-F., Diethlem, K.: Fractional-order attractors synthesis via parameter switchings. Commun. Nonlinear Sci. \textbf{15}(12) 3745-3753, (2011)

	\bibitem{md} Danca, M.-F., Garrappa R., Tang, W.K.S., Chen, G.: Sustaining stable dynamics of a fractional-order chaotic financial system by parameter switching. Comput. Math. Appl. \textbf{66}(5), 702-716 (2013)

	\bibitem{moha} Faieghi, M.R., Delavari, H.: Chaos in fractional-order Genesio-Tesi system and its synchronization. Commun. Nonlinear. Sci. \textbf{17}(2), 731-741 (2012)

	\bibitem{tava4} Tavazoei, M.S., Haeri, M.: A proof for non existence of periodic solutions in time invariant fractional order systems. Automatica \textbf{45}(8), 1886-1890 (2009)

	\bibitem{eva} Kaslik, E., Sivasundaram, S.: Non-existence of periodic solutions in fractional-order dynamical systems and a remarkable difference between integer and fractional-order derivatives of periodic functions. Nonlinear Anal.-Real \textbf{13}(3), 1489-1497 (2012)

	\bibitem{Podlubny1999} Podlubny, I.: \emph{Fractional differential equations, }(Academic Press, San Diego, CA, 1999).

	\bibitem{Diethelm2010} K. Diethelm, \emph{The Analysis of Fractional Differential Equations, }(Springer-Verlag, Berlin 2010).

	\bibitem{KilbasSrivastavaTrujillo2006} Kilbas, A.A., Srivastava, H.M., Trujillo, J.J.: \emph{Theory and applications of fractional differential equations,} (North-Holland Mathematics Studies, Amsterdam, 2006).

	\bibitem{bale2} Baleanu, D., Diethelm, K., Scalas, E., Trujillo, J.J.: \emph{Fractional calculus models and numerical methods }(Series on Complexity, Nonlinearity and Chaos. World Scientific 2012).

	\bibitem{hart} Hartley, T.T., Lorenzo, C.F., Trigeassou, J.C., Maamri, N.: Equivalence of History-Function Based and Infinite-Dimensional-State Initializations for Fractional-Order Operators. J. Comput. Nonlin. Dyn. \textbf{8}(4), 041014 (2013)

	\bibitem{kai2} Diethelm, K.: An extension of the well-posedness concept for fractional differential equations of Caputo\textsc{\char13}s type. App. Anal. \textbf{93}, 2126 (2014)

	\bibitem{kai} Diethelm, K., Ford, N.J., Freed, A.D.: A Predictor-Corrector Approach for the Numerical Solution of Fractional Differential Equations. Nonlinear Dynam.\textbf{29}(1), 3-22 (2002).

	\bibitem{tava} Tavazoei, M.S., Haeri, M.: Chaotic attractors in incommensurate fractional order systems. Physica D \textbf{327}(20), 2628- 2637 (2008)

	\bibitem{tava3} Tavazoei, M.S., Haeri, M.: A necessary condition for double scroll attractor existence in fractional-order systems. Phys. Lett. A \textbf{367}(1-2), 102-113 (2007)

	\bibitem{leo2} Leonov, G.A., Kuznetsov, N.V., Mokaev, T.N.: Hidden attractor and homoclinic orbit in Lorenz-like system describing convective fluid motion in rotating cavity. Commun. Nonlinear. Sci. \textbf{28}, 166 (2015)

	\bibitem{rab1} Rabinovich, M.: Stochastic autooscillations and turbulence. Uspehi Physicheskih Nauk [in Russian] \textbf{125}(1), 123-168 (1978).

	\bibitem{rab2} Pikovski, A.S., Rabinovich, M.I., Trakhtengerts, V.I.: Onset of stochasticity in decay confinement of parametric instability. Sov. Phys. JETP \textbf{47}, 715-719 (1978).

\bibitem{rus1}Glukhovsky, A.B., Dolzhansky, F.V.: Three component models of convection in a rotating fluid. Izv. Acad. Sci. USSR, Atmos. Oceanic Phys. \textbf{16}, 311–318 (1980)
\bibitem{rus2}Denisov, G.G.: On the rigid body rotation in resisting medium, Izv. Akad. Nauk SSSR: Mekh. Tverd. Tela \textbf{4}, 37–43 (1989) (in Russian)
\bibitem{rus3}Glukhovsky, A.B.: Nonlinear systems that are superpositions of gyrostats. Sov. Phys. Dokl. \textbf{27}(10), 823–825 (1982)
\bibitem{rus4}Dovzhenko, V.A., Dolzhansky, F.V.: Generating of the vortices in shear flows. Theory and experiment. Nauka, Moscow, 1987, (in Russian)	

\bibitem{dan} Danca, M.-F., Feckan, M., Kuznetsov, N., Chen, G.: Looking More Closely at the Rabinovich-Fabrikant System. Int. J. Bifurcat. Chaos \textbf{26}(2), 1650038 (2016)

	\bibitem{dan2} Danca, M.-F., Kuznetsov, N., Chen, G.:  Unusual dynamics and hidden attractors of the Rabinovich-Fabrikant system. submitted to Nonlinear Dynam.

	\bibitem{nickx} Kuznetsov, N.V.: Hidden attractors in fundamental problems and engineering models, arXiv:1510.04803v1 [nlin.CD]

	\bibitem{kuzbb} Kuznetsov, N., Kuznetsov, O., Leonov, G., Vagaitsev, V.: Analytical-numerical method for attractor localization of generalized Chua's system. IFAC Proceedings Volumes. \textbf{43}(11), 29-33 (2010)



\end{thebibliography}
\end{document}